\documentclass[
 aip,
 amsmath,amssymb,floatfix,
 reprint,%
]{revtex4-1}

\usepackage{graphicx}
\usepackage{dcolumn}
\usepackage{bm}

\usepackage[utf8]{inputenc}
\usepackage[T1]{fontenc}
\usepackage{mathptmx}
\usepackage{etoolbox}

\makeatletter
\def\@email#1#2{%
 \endgroup
 \patchcmd{\titleblock@produce}
  {\frontmatter@RRAPformat}
  {\frontmatter@RRAPformat{\produce@RRAP{*#1\href{mailto:#2}{#2}}}\frontmatter@RRAPformat}
  {}{}
}%
\makeatother
\begin{document}

\preprint{AIP/123-QED}

\title{Multimode fiber based high-dimensional light analyzer}
\author{Yuxuan Xiong}
 \affiliation
{School of Optical and Electronic Information, Wuhan National Laboratory for Optoelectronics, Next Generation Internet Access National Engineering Laboratory, and Hubei Optics Valley Laboratory, Huazhong University of Science and Technology, Wuhan, China}
\author{Hao Wu*}%
  \affiliation{School of Optical and Electronic Information, Wuhan National Laboratory for Optoelectronics, Next Generation Internet Access National Engineering Laboratory, and Hubei Optics Valley Laboratory, Huazhong University of Science and Technology, Wuhan, China}
\author{Mingming Zhang}
\affiliation
{School of Optical and Electronic Information, Wuhan National Laboratory for Optoelectronics, Next Generation Internet Access National Engineering Laboratory, and Hubei Optics Valley Laboratory, Huazhong University of Science and Technology, Wuhan, China}
\author{Yucheng Yao}
\affiliation
{School of Optical and Electronic Information, Wuhan National Laboratory for Optoelectronics, Next Generation Internet Access National Engineering Laboratory, and Hubei Optics Valley Laboratory, Huazhong University of Science and Technology, Wuhan, China}
\author{Ming Tang*}
\affiliation
{School of Optical and Electronic Information, Wuhan National Laboratory for Optoelectronics, Next Generation Internet Access National Engineering Laboratory, and Hubei Optics Valley Laboratory, Huazhong University of Science and Technology, Wuhan, China}
\email{wuhaoboom@hust.edu.cn, tangming@mail.hust.edu.cn}

\date{3 Feburary 2025}

\begin{abstract}
The wavelength and state of polarization (SOP) are fundamental properties of an optical field which are essential for applications in optical communications, imaging and other fields. However, it is challenging for existing spectrometers and polarimeters to measure these parameters simultaneously, resulting in reduced spatial and temporal efficiency. To overcome this limitation, we propose and demonstrate a compact multimode fiber (MMF)-based high-dimensional light analyzer capable of simultaneously performing high-precision measurements of both wavelength and SOP. Core-offset launching is introduced in the MMF to reshuffle the mode coupling. A neural network named WP-Net has been designed dedicated to wavelength and SOP synchronization measurements. Physics-informed loss function based on optical prior knowledge is used to optimize the learning process. These advancements have enhanced the sensitivity, achieving a wavelength resolution of 0.045 pm and an SOP resolution of 0.0088. 
\end{abstract}

\maketitle

\section{Introduction}
It is essential to measure both the wavelength and state of polarization (SOP) of light simultaneously. This comprehensive understanding of the properties of light facilitates a wide range of critical applications, including optical communication\cite{awajiReviewSpacedivisionMultiplexing2019}, remote sensing\cite{iglesiasInstrumentationSolarSpectropolarimetry2019}, chemical and biological studies\cite{liuMultimodeMicrofiberSpecklegram2024}, and astronomical observation\cite{glenarAcoustoopticImagingSpectropolarimetry}. Despite significant advancements in the design of polarimeters and spectrometers, it remains challenging to capture high-dimensional information including polarization and wavelength simultaneously. Modern instruments are typically limited to measuring only wavelength or polarization. Recent research has demonstrated the potential for simultaneous wavelength and SOP measurement using tunable liquid crystal metasurfaces\cite{niComputationalSpectropolarimetryTunable2022}. However, the spectral range is limited to near-infrared light, and the reflective design introduces additional complexity of the system. Some researchers have used simple thin-film interfaces with spatial and frequency dispersion to synchronously measure wavelength and polarization. This miniaturized system has a simple structure, but the accuracy of polarization state identification is relatively low\cite{fanDispersionassistedHighdimensionalPhotodetector2024}.Other researchers have constructed a multiplexing system that can simultaneously measure wavelength and SOP by multiplexing time-division and space-division methods \cite{stamSpectropolarimeterPlanetaryExploration2017,sterzikBiosignaturesRevealedSpectropolarimetry2012}. Inevitably, the multiplexing leads to a further increase in the shape factor and complexity of the system. Therefore, it is indispensable to propose a simple-structure and high-precision system capable of simultaneous wavelength and SOP measurement. 

The MMF, which is cost-effective and sensitive to the characteristics of the input light, has exhibited the potential to be a solution for achieving simultaneous wavelength and SOP measurements. This functionality is realized through MMF mode coupling, with a camera serving as the detector to capture the resulting speckle \cite{liuDeepLearningBasedSimultaneous2024,gaoSpatiallyresolvedBendingRecognition2023}. Researchers utilize techniques based on speckle imaging and advanced data analysis methods, including image differencing\cite{fujiwaraEvaluationImageMatching2018,reisStructuralHealthMonitoring2017}, zero-normalized cross-correlation\cite{caiReflectiveTactileSensor2023,liuReflectiveOpticalTactile2023}, and normalized inner product coefficients\cite{yuSubmicrometerDisplacementSensing1993}, to quantitatively characterize the output speckles. There has been research conducted to demonstrate the sensitivity of MMF to wavelength and SOP\cite{wangHighresolutionWavemeterBased2020,xiongMultimodeFiberSpeckle2024}. In recent research, the feasibility of measuring both wavelength and polarization simultaneously using MMF with the transmission matrix algorithm has been demonstrated. However, this approach still has potential for improvement due to the slow update rate of large-dimension matrices, especially when measuring two parameters simultaneously\cite{zhouHighaccuracySimultaneousMeasurement2024}. Consequently, a more efficient analytical approach is essential. Deep learning techniques have been proved to be highly effective for analyzing speckle images. Speckle images are utilized to train models and correlations can be established between intricate variations in the speckles and the target measurements. Researchers have already employed deep learning in combination with MMF to achieve sensing measurements of parameters, such as temperature, strain and tactile response\cite{liuDeepLearningBasedSimultaneous2024,kangDeepLearningEnabledTwoDirectional2024,ding2DTactileSensor2021,wangUltrasensitiveFiberEndTactile2023,sarkarDynamicMultimodeFiber}. In our previous work, we have demonstrated the use of convolutional neural network (CNN) for high-precision SOP measurements, with the system trained to be robust to environmental perturbations\cite{xiongMultimodeFiberSpeckle2024}.

\begin{figure*}
\includegraphics[width=\textwidth]{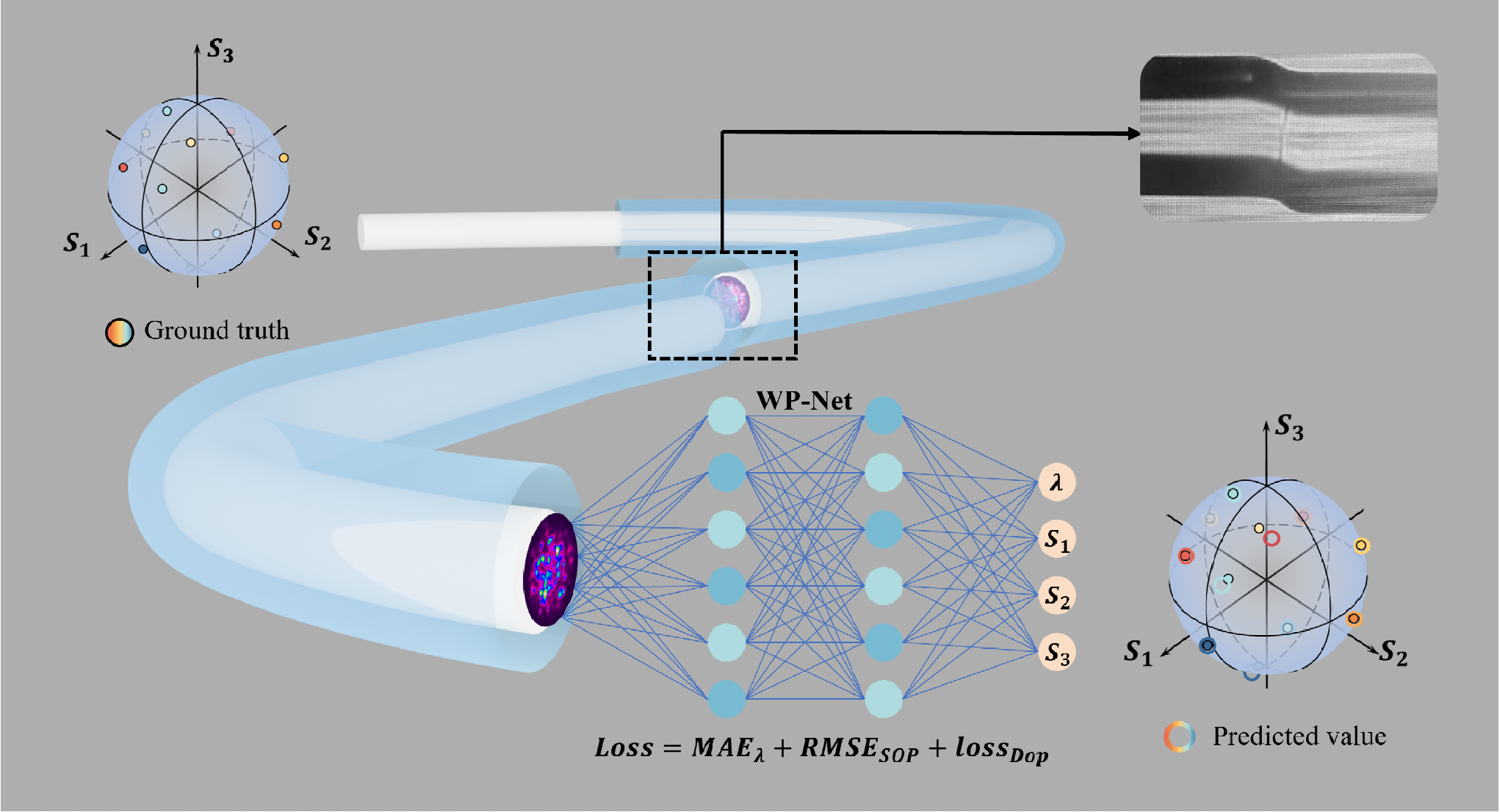}
\caption{\label{fig1} Scheme of the high-dimensional analyzer based on MMF. The introduction of core-offset launching and WP-Net has improved the sensitivity and accuracy.}
\end{figure*}

In this work, we propose a high-dimensional light analyzer based on MMF, as illustrated in Fig.~\ref{fig1}. Light with random wavelength and SOP generates corresponding speckle patterns at the output end of MMF. Accurate multi-parameter measurement is achieved by establishing a mapping relationship via a CNN. To enhance sensitivity, the intermediate mode reshuffle is introduced through core-offset launching, as shown in the dotted box in Fig.~\ref{fig1}. Additionally, a physical principle is integrated into the loss function, which improves the general applicability of the model and results in higher accuracy. We refer to this neural network for simultaneous wavelength and polarization measurement as WP-Net. Experimental results demonstrate high-precision simultaneous measurements of wavelength and SOP, with resolutions of 0.045 pm and 0.0088, respectively.

\section{Principle}
The large diameter of the core of an MMF enables the transmission of a multitude of modes. Different modes transmit across MMF with various speeds, resulting in a phase difference that produces a bright and dark mode interference pattern at the end facet of MMF, as illustrated in Eq.~\ref{eq:one}\cite{takaiStatisticalPropertiesLaser1985}. In this equation, $I$ is the speckle image at the end facet of MMF. $N$ denotes the number of modes. The amplitude and polarization of the $i$th mode is represented by $a_i$ and ${{\hat e}_i}(x,y)$, respectively\cite{rejaMultimodeOpticalFiber2024}. A change in the wavelength or SOP of the incident light results in a differential alteration of the optical range length of each mode. This causes a shift in the relative phase of the modes, resulting in a modification of the speckle pattern. Consequently, MMFs are capable of effectively measuring the optical field characteristics of the incident light, as variations in wavelength and SOP are directly reflected in changes to the speckles. 

\begin{equation}
I{(x,y,L)} = {\left| {\sum\limits_{i = 1}^N {{a_i}{{\hat e}_i}(x,y)\exp (j\frac{{2\pi }}{\lambda }{n_{eff,i}}L)} } \right|^2}
\label{eq:one}
\end{equation}

Similarity is used to evaluate the sensitivity of MMF-based measurement systems\cite{reddingAllfiberSpectrometerBased2013}, as shown in Eq.~\ref{eq:two}. In Eq.~\ref{eq:two}, $I$ and $I'$ are the two speckle images before and after a change in $\alpha$, where $\alpha$ is the parameter to be measured. $x$ is position on the speckle, ${\left\langle {} \right\rangle _x}$ is averaging over the speckle image and $\sigma$ is the standard deviation of the speckle image. The similarity does not directly determine the accuracy limit. When measuring a single parameter, the accuracy is defined by the smallest detectable variation in the speckle. The speckle correlation limit indicates the minimum distance at which neighboring parameter values can still be distinguished. A faster decline in the similarity curve over the same measurement range suggests a faster decorrelation of speckle patterns, which in turn reflects a higher resolution capability\cite{facchinUnveilingSignificanceIntrinsic2024}. Long MMF system exhibits faster decorrelation rate between speckles, leading to better resolution compared to short MMF system. \cite{reddingAllfiberSpectrometerBased2013}
\begin{equation}
S(\Delta \alpha ) = {\left\langle {\left( {\frac{{I - {{\left\langle I \right\rangle }_x}}}{{{\sigma _I}}}} \right)\left( {\frac{{I' - {{\left\langle {I'} \right\rangle }_x}}}{{{\sigma _{I'}}}}} \right)} \right\rangle _x}
\label{eq:two}
\end{equation}

\section{Methods}
\begin{figure}
  \includegraphics[height=3.5cm, keepaspectratio]{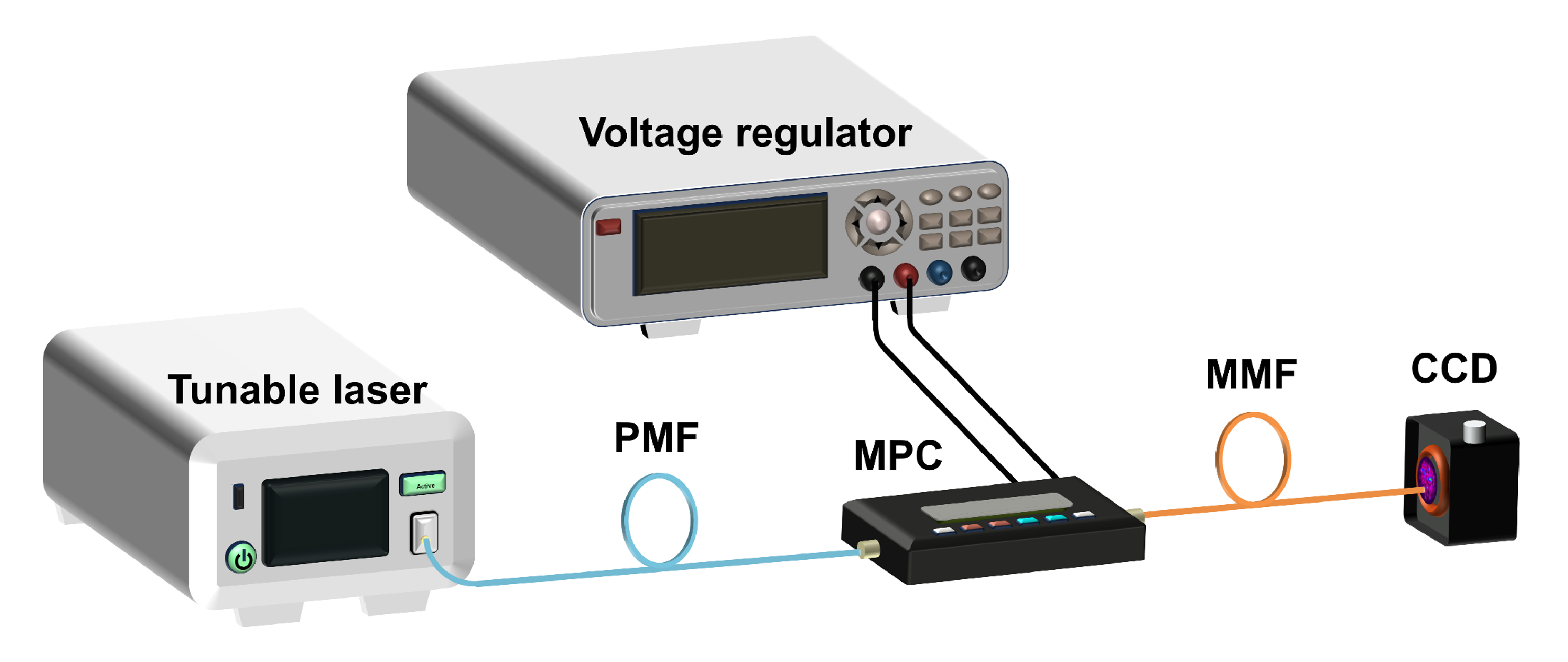}
  \caption{Schematic of the experimental setup}
  \label{fig2}
\end{figure}

Fig.~\ref{fig2} schematically illustrates the measurement system. Tunable laser (TSL 570, Santec) emits spectrally known light which passes through a motorized polarization controller (MPC, fiberpro, containing two independent rotating 1/4 waveplates) to produce an output in any SOP. This output light enters a MMF (2m length, SI 105/125-22/250, YOFC) as the light to be measured, excites mode coupling and forms a speckle at the end facet of the MMF, which is detected by a charge coupled device (CCD, SP620U-1550, Spiricon). The original image captured by the CCD has a resolution of 1200×1600. Only the portion containing the complete scattering is cropped out, with a size of 600×600. To shorten the learning time, we preprocessed the image with a kernel of 3×3 for maximum pooling, reducing the size to 200×200.

\begin{figure}
  \includegraphics[width=0.52\textwidth]{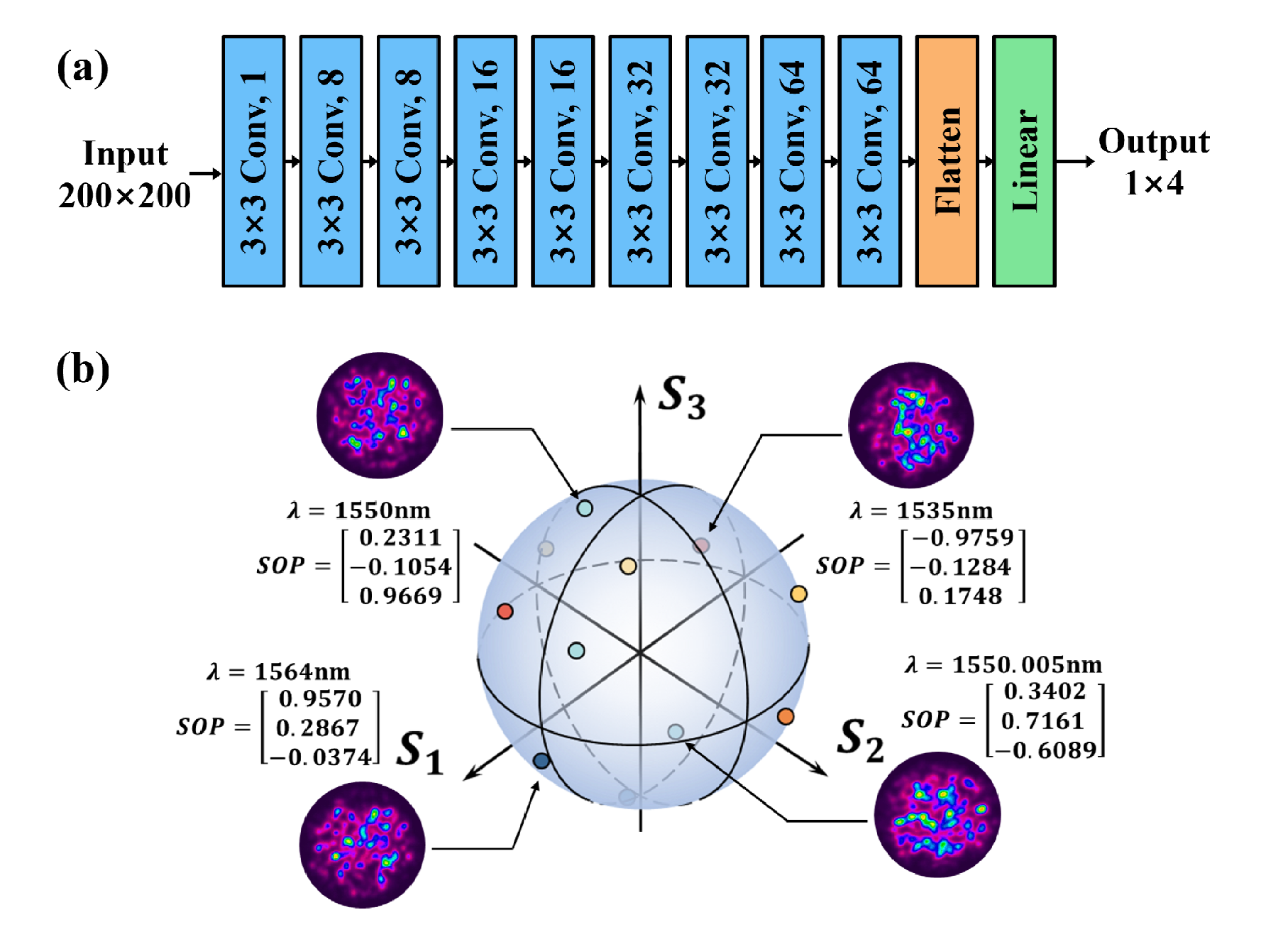}
  \caption{(a) Structure of the CNN to establish the mapping relationship between speckle, wavelength and SOP; (b) Each speckle corresponds to different wavelength and SOP.}
  \label{fig3}
\end{figure}

The cropped speckles are input into a neural network as shown in Fig~\ref{fig3}(a). Each speckle is labeled with its corresponding wavelength and Stokes parameters, depicted in Fig.~\ref{fig3}(b). The performance of CNN is evaluated using two metrics: the mean absolute error (MAE) for wavelength prediction, shown in Eq.~\ref{eq:three}, and the root-mean square error (RMSE) for SOP prediction, shown in Eq.~\ref{eq:four}. The MAE quantifies the accuracy of wavelength prediction as the absolute difference between the predicted and actual wavelengths, while the RMSE measures the accuracy of SOP prediction as the vector difference between the predicted and actual Stokes parameters. The loss function of the network is defined as the sum of these two-error metrics, as expressed in Eq.~\ref{eq:five}. The initial value of the learning rate is set at 0.0003, which gradually decreases as the learning process progresses. The training process is terminated if the loss does not decrease after 600 iterations.

\begin{equation}
MA{E_\lambda } = \left| {{\lambda _{{\rm{pre}}}} - {\lambda _{act}}} \right|
\label{eq:three}
\end{equation}

\begin{equation}
RMS{E_{SOP}} = \sqrt {{{({S_{{\rm{1 pre}}}} - {S_{{\rm{1 act}}}})}^2} + {{({S_{{\rm{2 pre}}}} - {S_{{\rm{2 act}}}})}^2} + {{({S_{{\rm{3 pre}}}} - {S_{{\rm{3 act}}}})}^2}}
\label{eq:four}
\end{equation}

\begin{equation}
L{\rm{oss}} = MA{E_\lambda } + RMS{E_{SOP}}
\label{eq:five}
\end{equation}

\section{Results and discussion}

\subsection{Wavelength and SOP analyzer}
\begin{figure*}
  \includegraphics[width=0.8\textwidth]{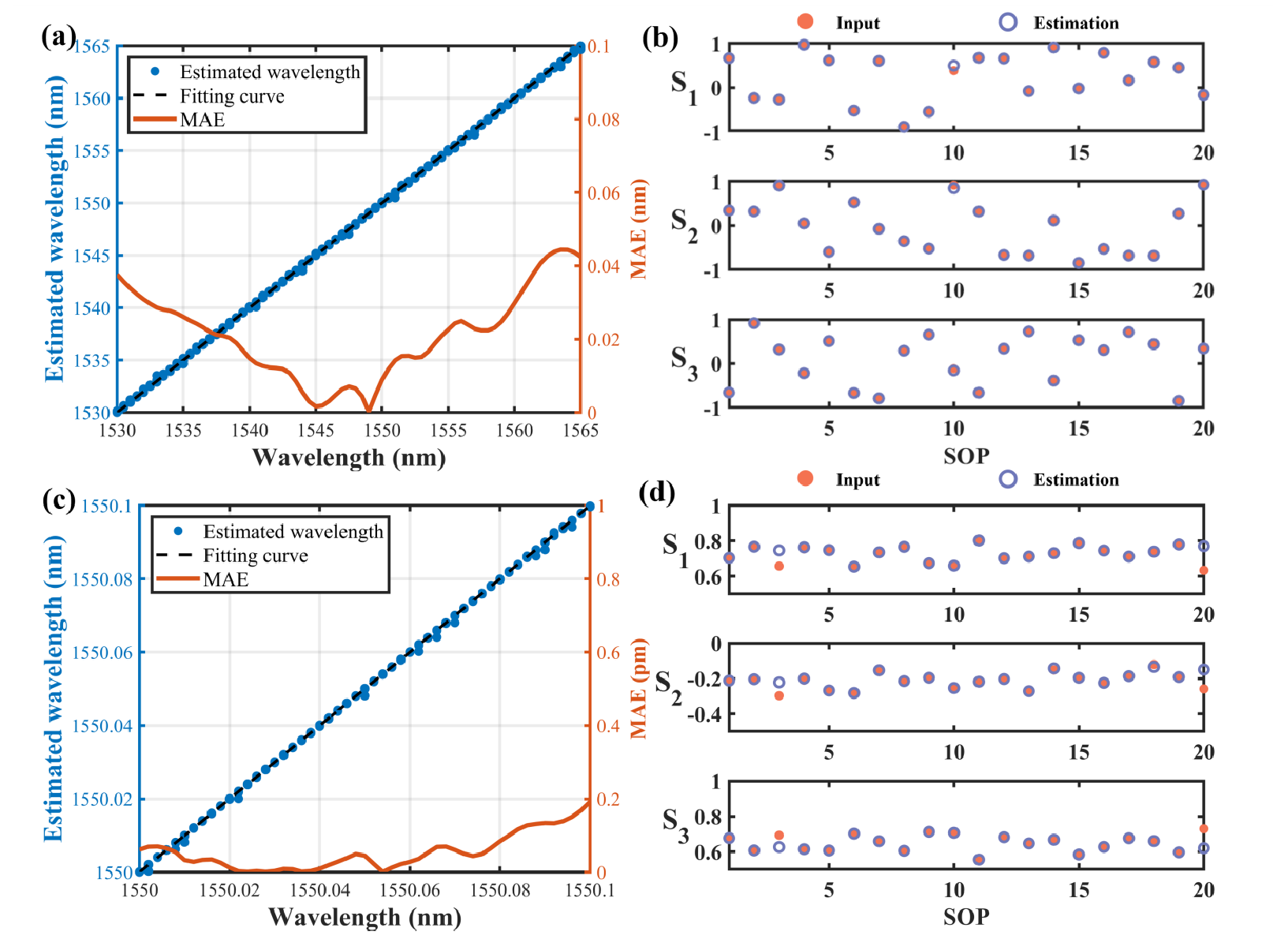}
  \caption{\label{fig4}The resolutions of the MMF-based analyzer for different parameter intervals. (a)-(b) Wavelength and SOP in the large-step dataset; (c)-(d) Wavelength and SOP in the small-step dataset.}
\end{figure*}

To address the trade-off between range and accuracy in the MMF speckle measurement system, high-precision measurement is achieved by integrating models with two measurement ranges and intervals. Two datasets were collected. The large-step dataset to assess the performance across a wide bandwidth consists of 28,400 speckles, covering a bandwidth of 35 nm with an interval of 0.5 nm. For each wavelength, 400 speckles are acquired corresponding to different SOPs covering the entire Poincare sphere. The average vector distance of the adjacent SOP vectors on the Poincare sphere in the large-step data set is 0.1856. The other dataset is the small-step dataset which aims to determine the accuracy limit of the system. It contains 20,400 speckles, with a bandwidth of 0.1 nm and an interval of 2 pm. Similarly, 400 speckles corresponding to different SOPs are acquired for each wavelength. The average vector distance of the adjacent SOPs in the small-step dataset is 0.0115. Both datasets were divided into training, validation, and test sets in an 8:1:1 ratio. The results are summarized in Table ~\ref{tb:one} and illustrated in Fig.~\ref{fig4}.

\begin{table}
\caption{\label{tb:one}The accuracy of the MMF-based analyzer at different parameter intervals}
\begin{ruledtabular}
\begin{tabular}{lll}
    Dataset  & Large-step & small-step \\
    \hline
    Bandwidth & 35 nm & 100 pm  \\
    Wavelength interval & 0.5 nm & 2 pm  \\
    Wavelength error (MAE)  & 0.026 nm & 0.085 pm  \\
    SOP interval & 0.1856 & 0.0115 \\
    SOP error (RMSE) & 0.1606 & 0.0138 \\
    $S_1$ (MAE) & 0.0947 & 0.0079 \\
    $S_2$ (MAE) & 0.0983 & 0.0086 \\
    $S_3$ (MAE) & 0.0846 & 0.0081 \\
\end{tabular}
\end{ruledtabular}
\end{table}

The system has been demonstrated to provide highly accurate wavelength measurements across different bandwidths. Notably, the wavelength error in the large-step dataset (26 pm) is smaller than the bandwidth of the small-step dataset (100 pm). This suggests that, when the two datasets are combined, the system can offer both a wide range and high accuracy in wavelength measurements. In the large-step dataset, the system can cover a broad range of wavelength and SOP measurements. However, in the small-step dataset, the system can only provide accurate wavelength measurements, with significant errors in the SOP measurements, as shown in Fig.~\ref{fig4}(d). This limitation arises because the system relies on the MMF characteristics to extract input light field features and map them onto the speckle spatial distribution. As a result, it is challenging to distinguish the effects of wavelength and SOP by analyzing the spatial intensity distribution alone. Additionally, the resolution of MMF-based system is positively correlated with the fiber length. However, increasing the fiber length not only raises the cost but also makes the system more sensitive to external environmental factors. To improve the feature extraction capability of short MMFs, we introduce mode reshuffle, which enhances the measurement accuracy of the system.

\subsection{Mode reshuffle}

Core-offset launching is introduced to reshuffle the mode coupling and enhance the sensitivity. To identify the accuracy improvement introduced by different displacements, 5 distinct displacement groups are established for comparison. These groups include: displacement 0, 1/8 fiber diameter, 1/4 fiber diameter, 3/8 fiber diameter, and 1/2 fiber diameter. The 0-displacement group was designed to examine the effect of the fusion splice point on accuracy.

\begin{figure}
  \includegraphics[width=0.48\textwidth]{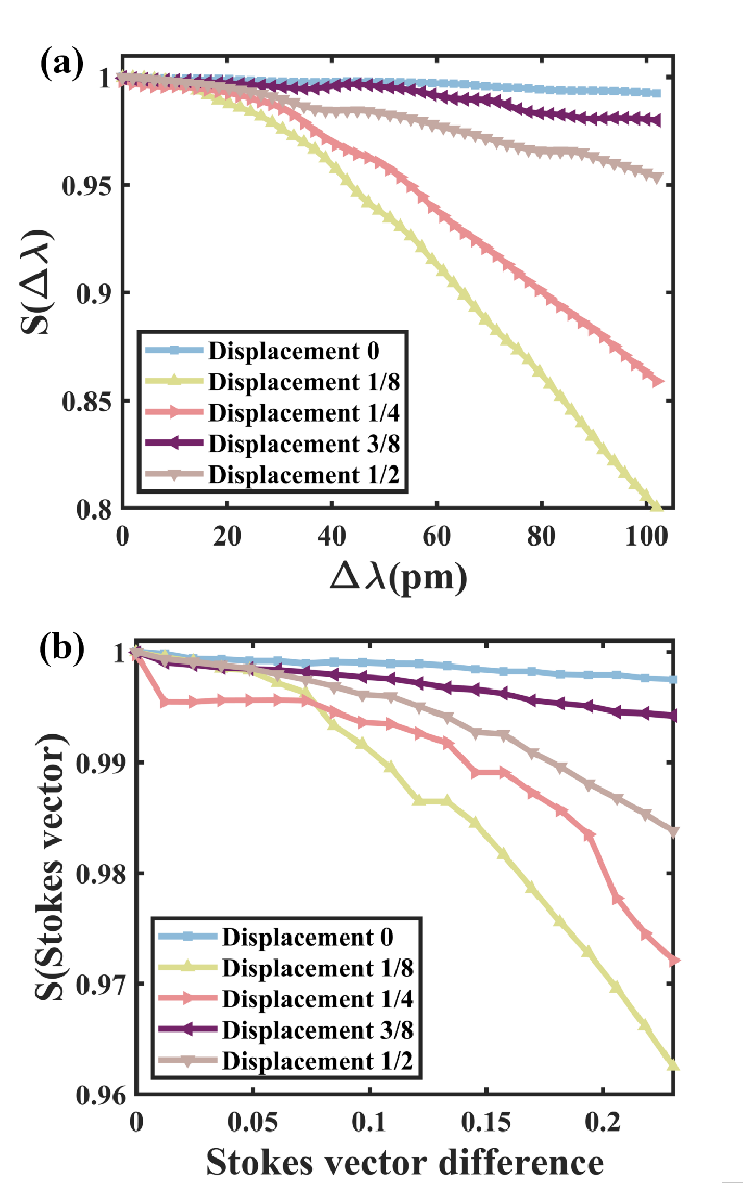}
  \caption{\label{fig5}Similarity curves of the variance to be measured corresponding to different displacements. (a) Wavelength; (b) SOP.}
\end{figure}

Similarity, as defined in Eq.~\ref{eq:two}, is utilized to quantify the sensitivity of the system. A change in the measured value leads to a reduction in the similarity between the images before and after the change. Within the same measurement range, a faster decorrelation rate corresponds to a higher system sensitivity. The wavelength and SOP similarity curves are shown in Fig.~\ref{fig5}. The mode reshuffle has significantly increased the decorrelation rate of wavelength and has slightly impact on the SOP. It is shown in Fig.~\ref{fig5} that the decorrelation rate does not follow a linear relationship with displacement. The 1/8 displacement demonstrates the fastest decorrelation rate and optimal mode reshuffle effects to enhance the sensitivity of both wavelength and SOP.

\begin{figure}
  \includegraphics[width=0.45\textwidth]{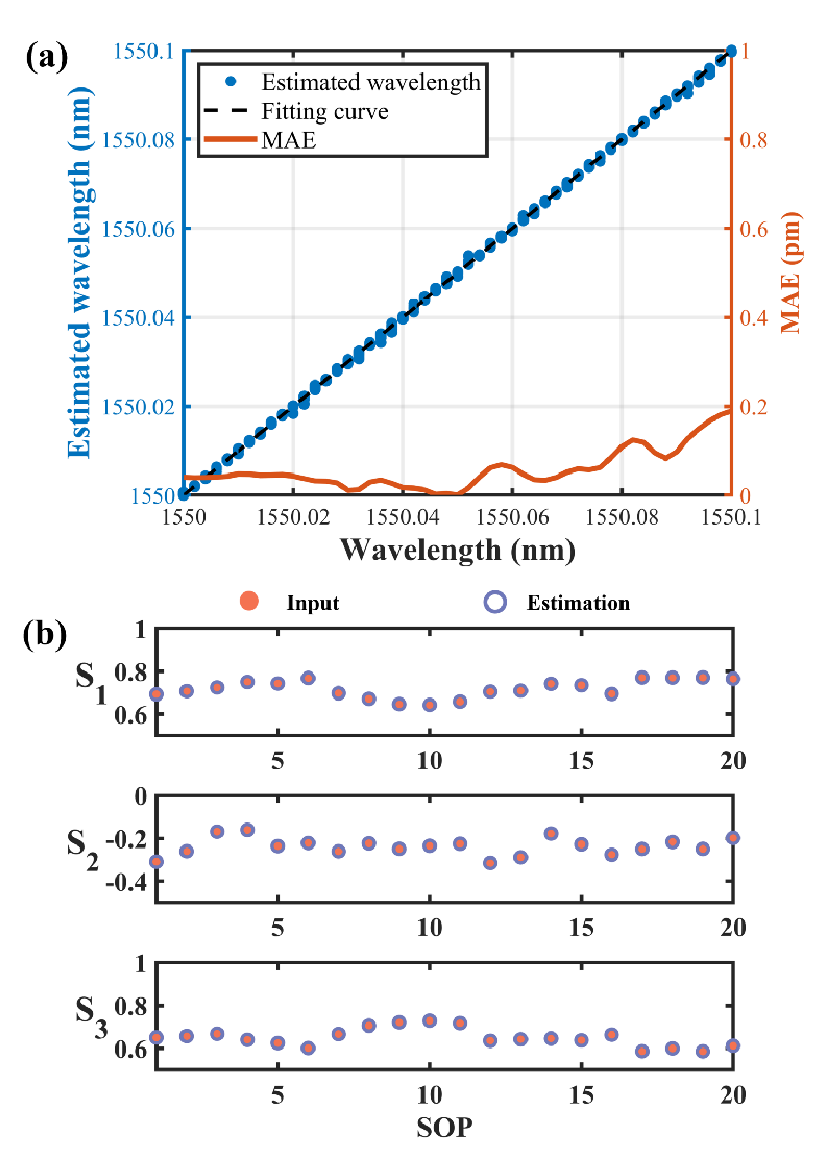}
  \caption{\label{fig6}The resolutions before and after introduction of mode reshuffle. (a) Wavelength; (b) SOP}
\end{figure}

Based on the above conclusion, the 1/8 displacement provides the greatest improvement in sensitivity. Therefore, the same dataset for the displacement of 1/8 was collected. The dataset contains 20,400 speckles with a bandwidth of 100 pm, an interval of 2 pm, with each wavelength corresponding to 400 different SOPs. The results are shown in Table ~\ref{tb:two}. Comparing with the original system, the wavelength resolution of the 1/8 displacement group reaches 0.065 pm. And the SOP resolution is reduced to 0.0103, smaller than the interval of SOP. There are improvements of 23.53\% and 25.36\% in wavelength and SOP respectively compared to the original system. The error distribution is shown in Fig.~\ref{fig6}.

\begin{table}
\caption{\label{tb:two}Comparison of system resolution before and after mode reshuffle}
\begin{ruledtabular}
\begin{tabular}{lll}
Displacement & Origin & 1/8 \\
    \hline
    Wavelength error (MAE)  & 0.085 pm & 0.065 pm  \\
    SOP error (RMSE) & 0.0138 & 0.0103 \\
    $S_1$ (MAE) & 0.0079 & 0.0059 \\
    $S_2$ (MAE) & 0.0086 & 0.0064 \\
    $S_3$ (MAE) & 0.0081 & 0.0045 \\
\end{tabular}
\end{ruledtabular}
\end{table}

\subsection{WP-Net with loss function optimization}

During the training process, it is notable that the 1/8 displacement case is more prone to overfitting. This is attributed to the higher decorrelation rate of the 1/8 displacement group, which results in increased sensitivity and more noticeable variations in the speckle compared to the original system. Therefore, we propose the WP-Net with additional physical constraint based on optical principles to the loss function. The WP-Net based on the optical principles aims to achieve simultaneous wavelength and SOP measurements with enhanced generalization capability. In this experiment, since fully polarized light, with a polarization degree of 1, is used in this experiment, the Stokes parameters satisfy the relationship of ${S_0}^2$ =$ {S_1}^2$ + ${S_2}^2$ + ${S_3}^2$.This optical prior knowledge can be integrated into the loss function, as shown in Eq.~\ref{eq:six}. The modified loss function is expressed in Eq.~\ref{eq:seven}.
\begin{equation}
los{s_{DOP}} = \frac{{S_1^2 + S_2^2 + S_3^2}}{{S_0^2}} - 1
\label{eq:six}
\end{equation}

\begin{equation}
L{\rm{oss'}} = MA{E_\lambda } + RMS{E_{SOP}} + los{s_{DOP}}
\label{eq:seven}
\end{equation}

\begin{figure}
  \includegraphics[width=0.48\textwidth]{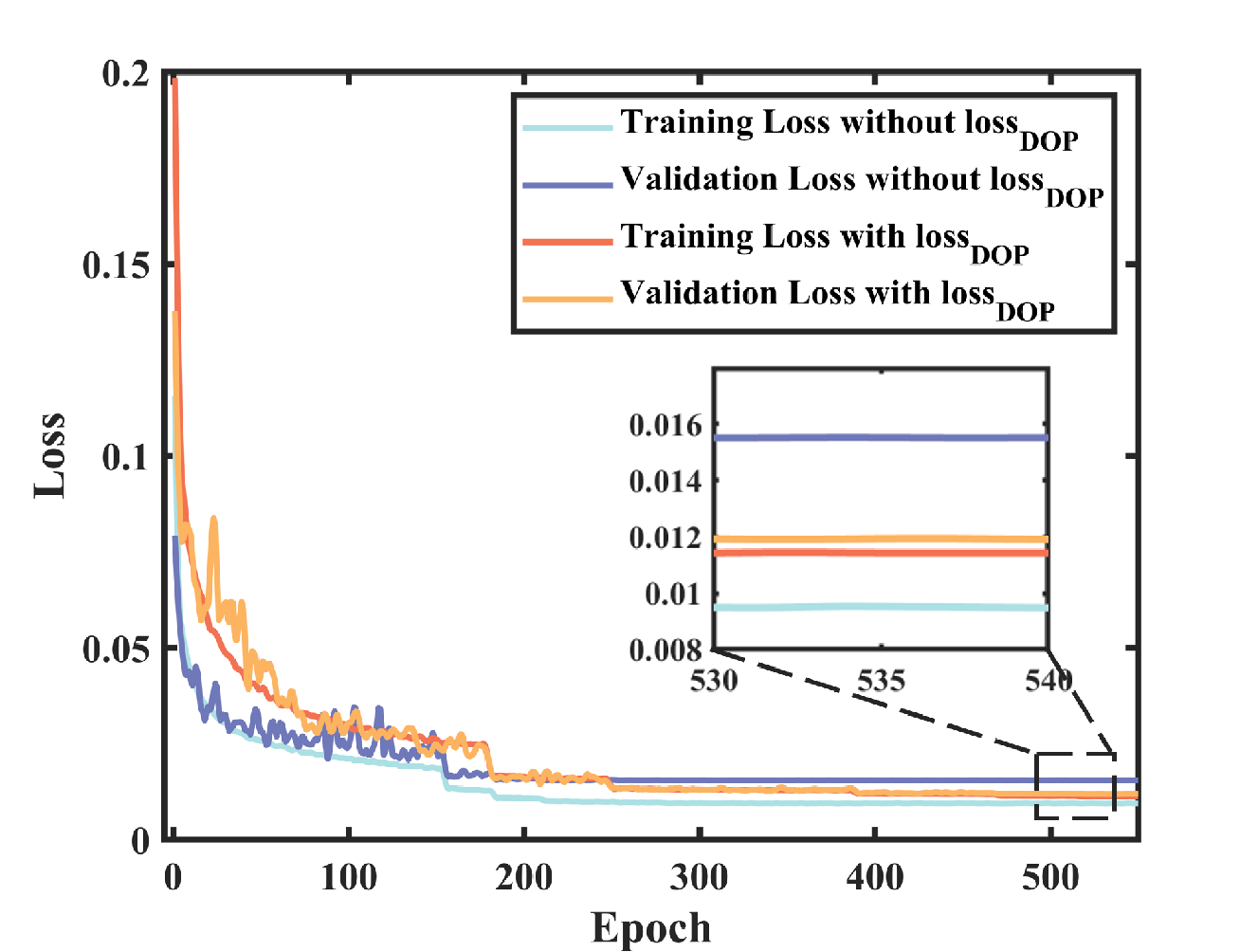}
  \caption{\label{fig7}The comparison of loss curves with and without WP-Net}
\end{figure}

Training with WP-Net, the loss curve with and without additional physics-informed loss function are shown in Fig.~\ref{fig7}. Before using the WP-Net, the loss exhibited a little overfitting. However, with the implementation of WP-Net, the losses of training and validation decrease synchronously indicating an improvement in generalization capability. The accuracy results are shown in Table ~\ref{tb:three}. In summary, with the introduction of mode reshuffle and WP-Net, the resolution of wavelength and SOP improved significantly, from 0.085 pm and 0.0138 to 0.045 pm and 0.0088, respectively. This represents an enhancement of 47.06\% in wavelength resolution and 36.23\% in SOP resolution compared to the original system. 

\begin{table}
\caption{\label{tb:three}Wavelength and SOP resolution comparison between original system and system with mode reshuffle and WP-Net}
\begin{ruledtabular}
\begin{tabular}{lll}
Displacement & Origin & 1/8 with WP-Net \\
    \hline
    Wavelength error (MAE)  & 0.085 pm & 0.045 pm  \\
    SOP error (RMSE) & 0.0138 & 0.0088 \\
    $S_1$ (MAE) & 0.0079 & 0.0057 \\
    $S_2$ (MAE) & 0.0086 & 0.0056 \\
    $S_3$ (MAE) & 0.0081 & 0.0044 \\
\end{tabular}
\end{ruledtabular}
\end{table}

\section{Conclusion}

We proposed an MMF-based analyzer for high-dimensional light achieving precise wavelength and SOP measurements. Mode reshuffle introduced by core-offset launching enhances short MMF system sensitivity. A WP-Net with physics-informed loss function based on the optical knowledge is designed to accelerate network convergence and augment prediction precision. The system merges two models covering distinct measurement ranges, enabling simultaneous detection of wavelength and SOP across an extensive spectral range with elevated accuracy. The MMF-based analyzer attains a wavelength resolution of 0.045 pm and an SOP resolution of 0.0088. It offers a compact, cost-effective, and spatiotemporally efficient solution suitable for optical communications, imaging, and other fields.

\begin{acknowledgments}
This work was supported by the National Natural Science Foundation of China under Grants 62225110, 61931010 and the Major Program (JD) of Hubei Province (2023BAA013).
\end{acknowledgments}

\section*{Conflict of interest}
The authors have no conflicts to disclose.

\section*{Data Availability Statement}
The data that support the findings of this study are available from the corresponding author upon reasonable request.

\nocite{*}
\bibliography{aipsamp}

\end{document}